\documentclass[aps,twocolumn]{revtex4-1}

\usepackage{bm}%

\usepackage[colorlinks=true,linkcolor=blue]{hyperref}%

\expandafter\ifx\csname package@font\endcsname\relax\else

 \expandafter\expandafter

 \expandafter\usepackage

 \expandafter\expandafter

 \expandafter{\csname package@font\endcsname}%

\fi

\usepackage{graphicx}
\usepackage{dcolumn}
\usepackage{bm}
\usepackage{amsmath}
\usepackage{amssymb}
\usepackage{pict2e}
\usepackage{multirow}
\usepackage{pifont}
\usepackage{color}

\newcommand{\rms}{\textit{rms}}
\newcommand{\pdf}{\textit{pdf}}

\begin{document}


\title{Influence of the impellers' geometry on turbulent von K\'arm\'an swirling flows}

\author{C.~Jause-Labert}
\email{clement.jause-labert@ec-lyon.fr}
\author{F.S.~Godeferd}
\affiliation{Laboratoire de M\'ecanique des Fluides et d'Acoustique UMR CNRS 5509,\\
\'Ecole centrale de Lyon, Universit\'e de Lyon,
 \'Ecully, France}

\date{\today}

\begin{abstract}
We report numerical evidences of the influence of the blades' shape and of their direction of rotation on turbulent flows generated by counter-rotating bladed-disks enclosed in a cylinder, known as the von K\'arm\'an swirling flow. We show that, although the mean flows generated by stirrers with straight or curved blades can be topologically similar, turbulence production 
occurs in different zones. The structure and intensity of the turbulent fluctuating field also varies with the geometry, 
and retains significant anisotropy. From the direct numerical simulations fields, we analyse the 
dynamics  of enstrophy and we  show that the distribution of
relative helicity $h=\bm{u}\cdot\bm{h}/(\vert \bm{u}\vert\vert\bm{\omega}\vert)$ 
depends on the forcing mecanism of turbulence.
\end{abstract}

\maketitle


\textit{Introduction} -- In order to experimentally study statistically stationary turbulence, scientists developed in the last two decades the so-called von K\'arm\'an device \cite{douady1991direct,cadot1995characterization,labbe96,raveletJFM}, designed
to produce inertial turbulent flows by the rotation of two facing bladed impellers enclosed in a cylindrical cell.  Different regimes for the turbulent flow are observed  when the rotation rates $f_1$  and $f_2$ of the two impellers are varied,
and for different aspect ratios $\Gamma=H/R$ of the fluid cell, where $H$ is the distance between the two disks and $R$ the inner radius of the cylindrical container. The kinematic viscosity of the fluid $\nu$ is also a parameter, and one builds a Reynolds number 
 $\mathrm{Re}_K=2\pi R^2 f\nu^{-1}$ based on the intensity of the forcing by the bladed disks $f=({f_1^2+f_2^2})^{1/2}$. However, the variability of the different experimental flows observed 
can only be explained by considering
 the precise geometry of the blades mounted on the rotating disks --- for example, a dynamo action can be produced with a certain kind of impellers \cite{monchaux2007generation} and not with other ones \cite{marie2003numerical} for an equivalent $\mathrm{Re}_K$. Many features of the turbulent flows observed with curved blades cannot be observed with straight ones: for instance, the multistability of the mean flow (between one- or two-cell states)  is observed with curved blades but nonexistent with straight blades \cite{raveletPRL}. Moreover, the direction of rotation of the blades, when curved, actually modify the properties of the flows \cite{monchauxPRL1,monchauxPRL2}. In addition, the 
 homogeneity and isotropy of turbulence in the core of the cell, once thought established, is now questioned. In this letter, we analyze the mechanisms of  turbulence production from the forcing by either straight or curved blades, the zones where energy
 is injected and transported, and how it influences the homogeneity and isotropy properties. Such an analysis is
 permitted by direct numerical simulations (DNS) with a volume-penalization technique for the rotating blades and the
 cylindrical container. We thus have access to the phenomena in a way not accessible by experimental realizations.

\begin{figure}
\begin{center}
\setlength{\unitlength}{0.01\textwidth}
\begin{picture}(48,19.5)
\put(0,0){\includegraphics[height=0.2\textwidth]{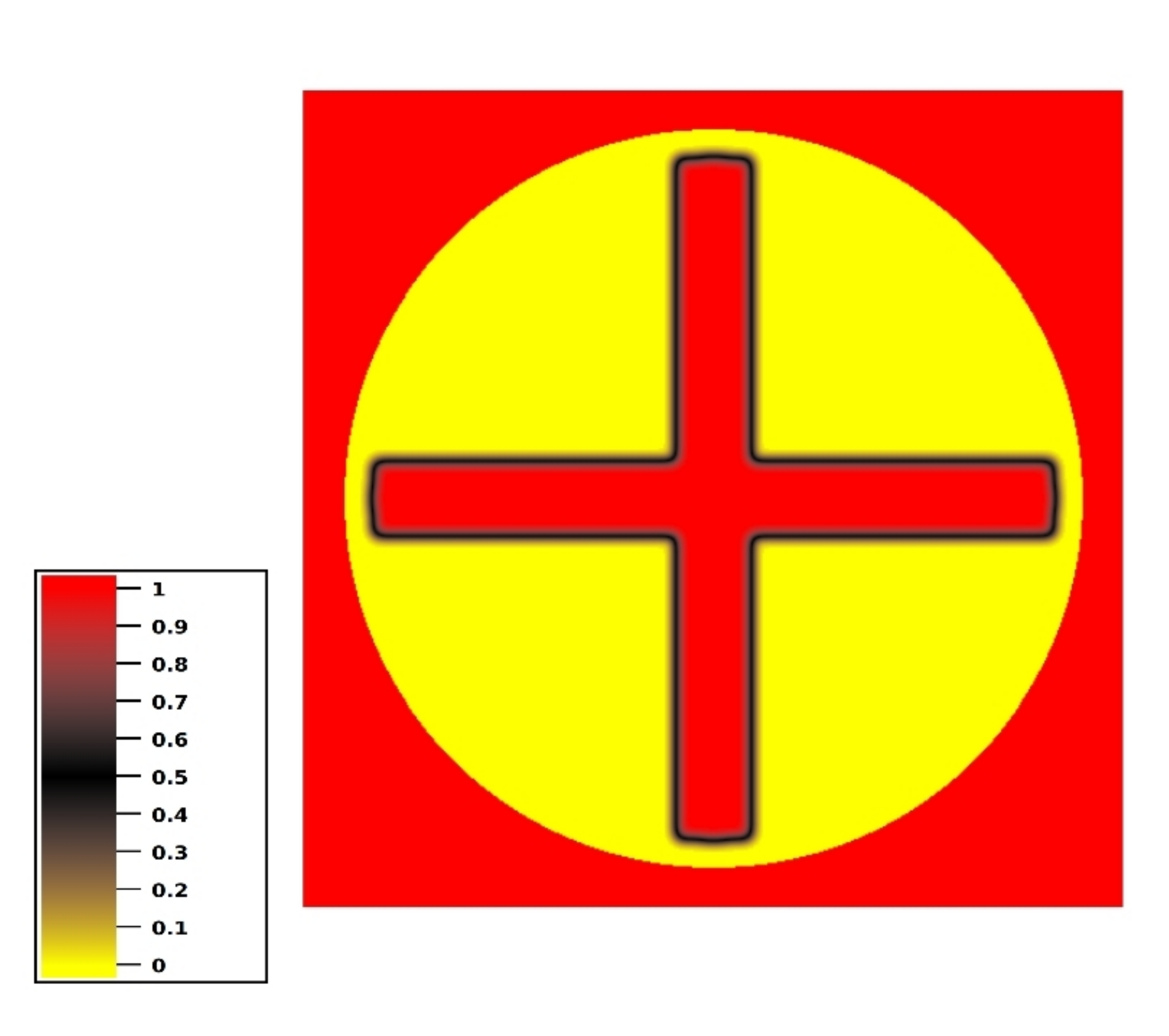}}
\put(24,0){\includegraphics[height=0.194\textwidth,bb=180 0 813.0375 720.6925,clip=]{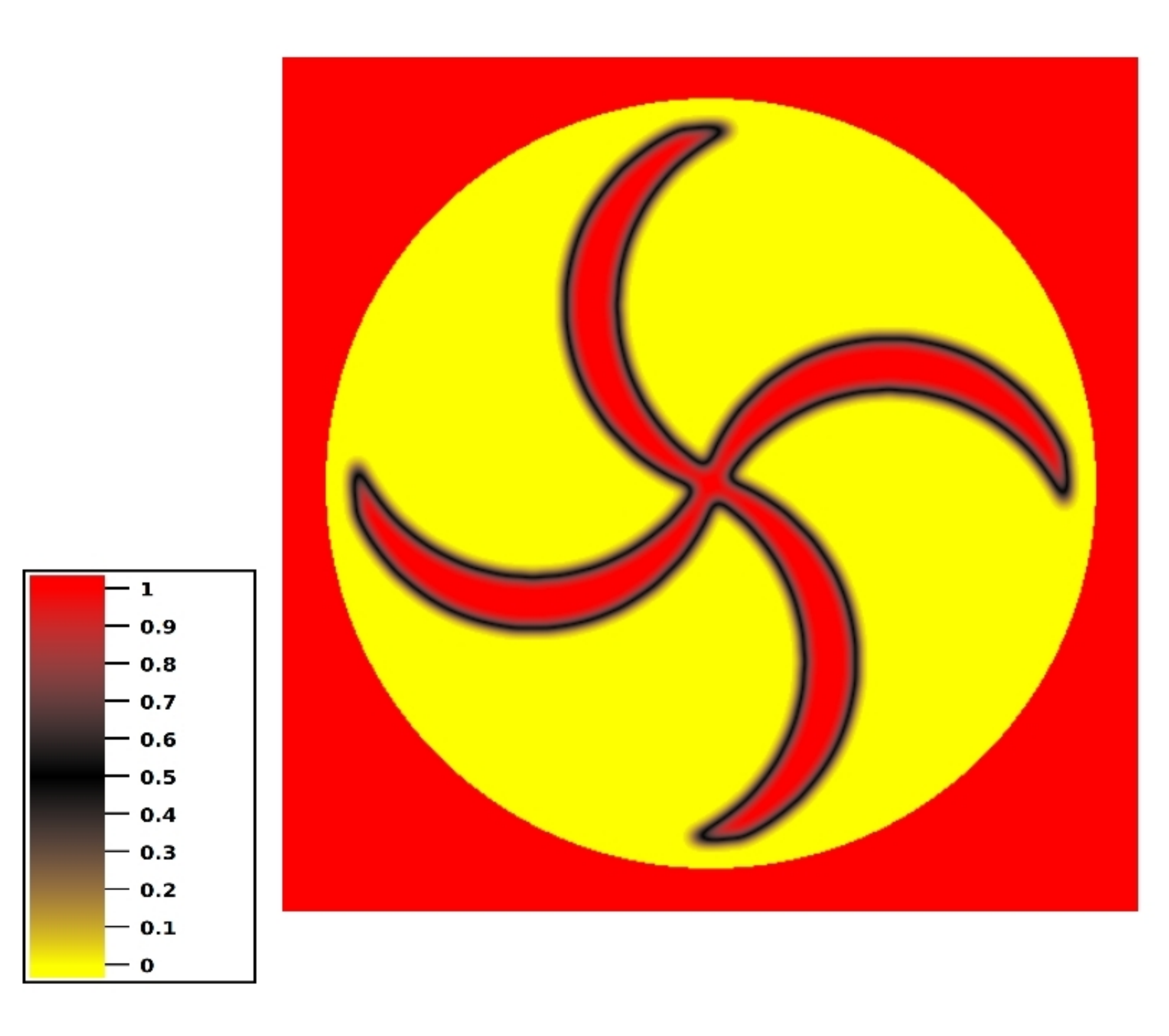}}

\thicklines
\qbezier(35,4.2)(38,5.5)(38.8,8.5)
\put(39.1,9.1){\vector(1,2.5){0}}
\put(37.5,3.5){\Large{$\bm{\oplus}$}}

\qbezier(34.5,16.7)(37.5,15.6)(38.5,12.8)
\put(38.7,12.3){\vector(1,-2.5){0}}
\put(37.5,15.5){\Large{$\bm{\ominus}$}}

\put(11.3,0){Case S}
\put(29.5,0){Case C$\pm$}

\end{picture}
\caption{\label{pales}(Color online) Geometries of the simulated blades, corresponding to the initial distribution of the mask function $\chi$ in a horizontal plane containing the blades (red=solid; yellow=fluid). For the curved blades, the direction of rotation is either negative $\ominus$ (concave face forward) or positive $\oplus$ (convex face forward).}
\end{center}
\end{figure}
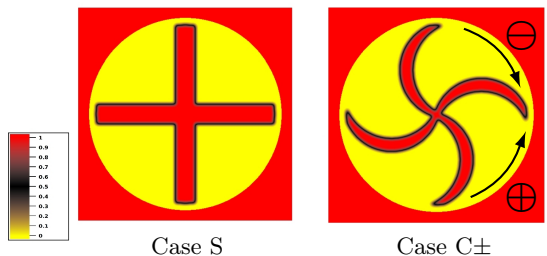

\textit{Flow geometry, parameters and algorithm} -- 
To allow comparison with existing experiments of von K\'arm\'an swirling flows, we consider a cylindrical
container in which two facing rotors are placed, with an aspect ratio of the fluid cell $\Gamma=1.8$, equivalent to \cite{raveletJFM,raveletPRL,monchauxPRL1,monchauxPRL2}. The radius $R_d$ of the disks is the same as the TM60 impellers used in these experimental works ($R_d=0.925R$) as well as the height of the blades ($H_b=0.2R$) and their curvature radius ($R_{curv}=0.5R$) for the curved ones. The main difference with the experiments is a number of blades reduced
to four (6 straight ones in \cite{cadot1995characterization}, 16 curved ones in \cite{raveletPRL} for example) in the simulations, due to numerical constraints, as shown on figure \ref{pales}.
We perform DNS of the Navier-Stokes equations  using a 3D Fourier pseudo-spectral method in a
periodic domain.  The above geometry of the flow and the no-slip boundary conditions are
imposed by a volume-penalization method \cite{angot}, which allows an accurate  representation of the fixed cylindrical container and of the mobile impellers. Technical details and validation of this method are explained in \cite{cafcjl}. 
The penalized Navier--Stokes equations solved are therefore
\begin{equation}
\frac{\partial\bm{u}}{\partial t}-\bm{u}\times\bm{\omega}=-\bm{\nabla}p_t+\nu\nabla^2\bm{u}-{\eta}^{-1}\chi\left(\bm{u}-\bm{u}_s\right),
\label{NSpen}
\end{equation}
where $\bm{u}$ and $\bm{\omega}=\bm{\nabla}\times\bm{u}$ are the velocity and vorticity fields, $p_t$ is the total pressure and $\chi$ is the mask function separating the spatial domain into solid ($\chi=1$)  and fluid ($\chi=0$) regions. The last term on the right-hand-side of equation \eqref{NSpen} is  the volume penalization term and the combined specification of $\chi(\bm{x},t)$ and of the
boundary velocity $\bm{u}_s(\bm{x},t)$ sets the required boundary conditions. $\eta$ is the penalization parameter chosen
as small as possible (typically $\eta\simeq 10^{-3}$).
In the present work, we focus on the exact counter-rotating regime $f_1=-f_2$, where $f=0.1$ with a fluid viscosity $\nu=2\times10^{-4}$ (all the parameters are non-dimensionalized). The radius of the cylinder is $R=0.9L_h$, where $L_h=2\pi$ is the horizontal size of the total resolution domain, whose vertical extent is $L_v=4\pi$. The Reynolds number is therefore $\mathrm{Re}_K\approx 25000$, 
above the value $20000$ at which fully developped turbulence is expected \cite{raveletJFM}, but well below that
for a bi-stable behavior, at $\mathrm{Re}_K\approx 10^{5}$ \cite{torreburguetePRL}. The number of collocation points  in the horizontal and vertical directions are $N_{x,y}=512$ and $N_z=1024$. We consider three different cases: case S with straight blades (left figure \ref{pales}) and case C- (resp. C+) with curved blades rotating in the negative (resp. positive) direction of rotation (right figure \ref{pales}). The stirrers' rotation is  progressively increased from rest to avoid a discontinuity in time.

\begin{figure}
\begin{center}
\setlength{\unitlength}{0.01\textwidth}
\begin{picture}(48,27)
\put(0,0){\includegraphics[height=44mm,bb=0 0 139 195.73125,clip=]{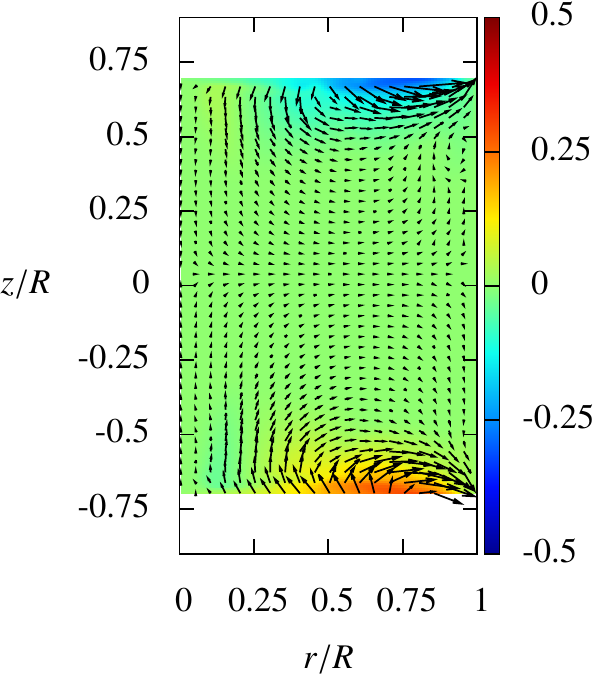}}
\put(31,0){\includegraphics[height=44mm,bb=46 0 171.64125 195.73125,clip=]{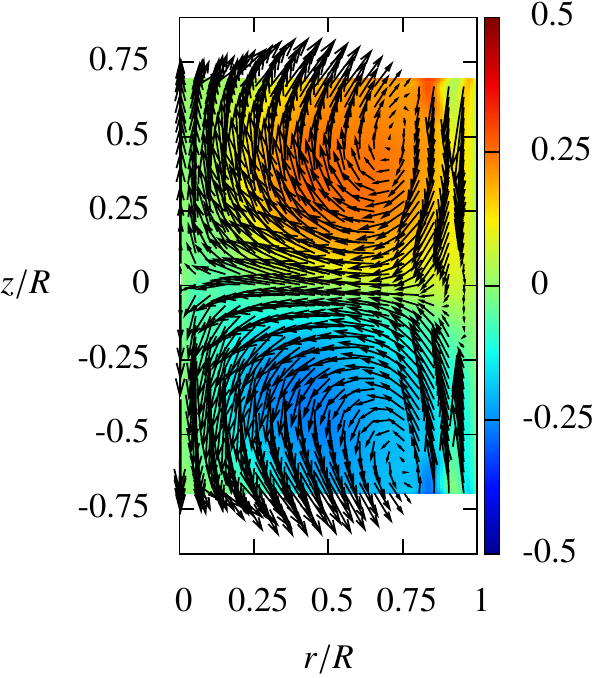}}
\put(18.4,0){\includegraphics[height=44mm,bb=46 0 139 195.73125,clip=]{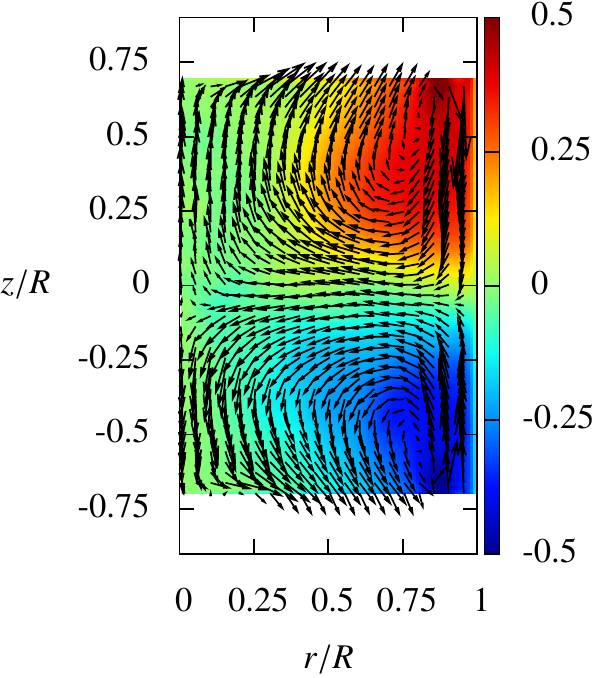}}
\put(8.6,25){Case C-}
\put(33.35,25){Case C+}
\put(21.8,25){Case S}

\put(28.2,22.5){$\otimes$}
\put(28.2,5.0){$\odot$}

\put(15.4,22.5){$\odot$}
\put(15.4,5.0){$\otimes$}

\put(40.8,22.5){$\otimes$}
\put(40.8,5.0){$\odot$}
\end{picture}
\caption{\label{mean}(Color online) Mean flows in a half vertical plane (symmetry axis on the left, solid wall on the right): C- for curved blades  in  negative direction, C+ for the positive direction, S for straight blades. In color the azimuthal velocity component $u_\theta$; the vectors are projections of the mean velocity field on the $rz$-plane.  $\mathrm{Re}_K\simeq25000$ for all cases. $\odot$ and $\otimes$ show the rotation direction  of the top and bottom impellers, located
in the white spaces.}
\end{center}
\end{figure}
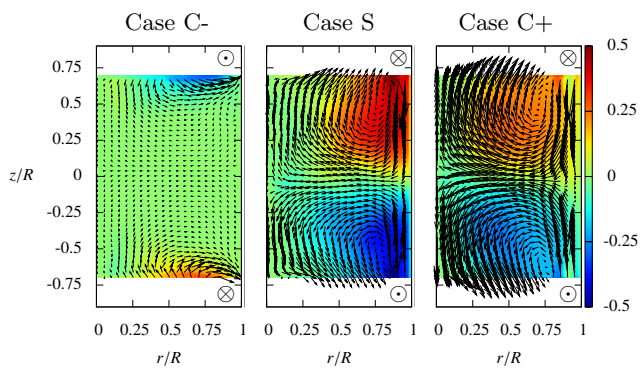

\textit{Statistically stationary states} -- Although at the same $\mathrm{Re}_K$, the different geometries 
induce important differences in the flow evolution and in the eventual statistically stationary state. First, the mean total kinetic energy in the fluid cell is $k=\left<\bm{u}^2\right>/2\simeq0.1$ for case S, but reduced to $k\simeq0.06$ for case C+ and to $k\simeq0.03$ for case C-. Consequently, the energy cascade from the large forced scales to the dissipative ones is reduced, so that the 
steady state is reached very shortly in case C- ($t_{stat}f\simeq0.5$),
but over a longer time in cases  S ($t_{stat}f\simeq3.5$) and C+ ($t_{stat}f\simeq6$). 
Figure \ref{mean} shows the corresponding computed mean flows, obtained by azimuthal and temporal averaging from $t=t_{stat}$ to the end of the computations.
Clearly,  the three mean flows are significantly different, especially for case C- where the curved blades rotating in this reverse direction are not able  to generate the same large pair of toroidal vortices in the fluid domain, as  cases S and C+, which  agree
with experimental observations for a symmetric forcing $f_1=-f_2$. Accordingly, case C- has too low energy to be considered turbulent --- the \rms~velocity in the core region is $({\left<u^2\right>_{core}})^{1/2}\approx10^{-4}$, compared to the maximum velocity of the impellers $u_{s,\mathrm{max}}=2\pi fR\approx 1.8$ ---  and is not discussed any more from here on.  
The two cases S and C+ exhibit noticeable differences: the magnitude of azimuthal velocity component is much larger 
for the straight blades case S than for the curved one C+. This dominance is however counter-balanced by larger vertical velocities towards the disks in the neighborhood of the impellers at radii $r^*\equiv r/R<0.5$ for case C+. In the curved blades simulation, one also notices a thin region $r^*>0.85$ which does not belong to the recirculation cells. In this region, when $r^*$ increases,  the azimuthal velocity component is depleted slightly before reaching a local maximum and connecting with
the inner boundary layer of the cylinder (wall at $r^*=1$). The C+  and S toroidal cells  therefore have a  different shape and size, and a different energy distribution.

\begin{figure}
\begin{center}
\setlength{\unitlength}{0.01\textwidth}
\begin{picture}(48,36)
\put(0,2){\includegraphics[height=0.33\textwidth]{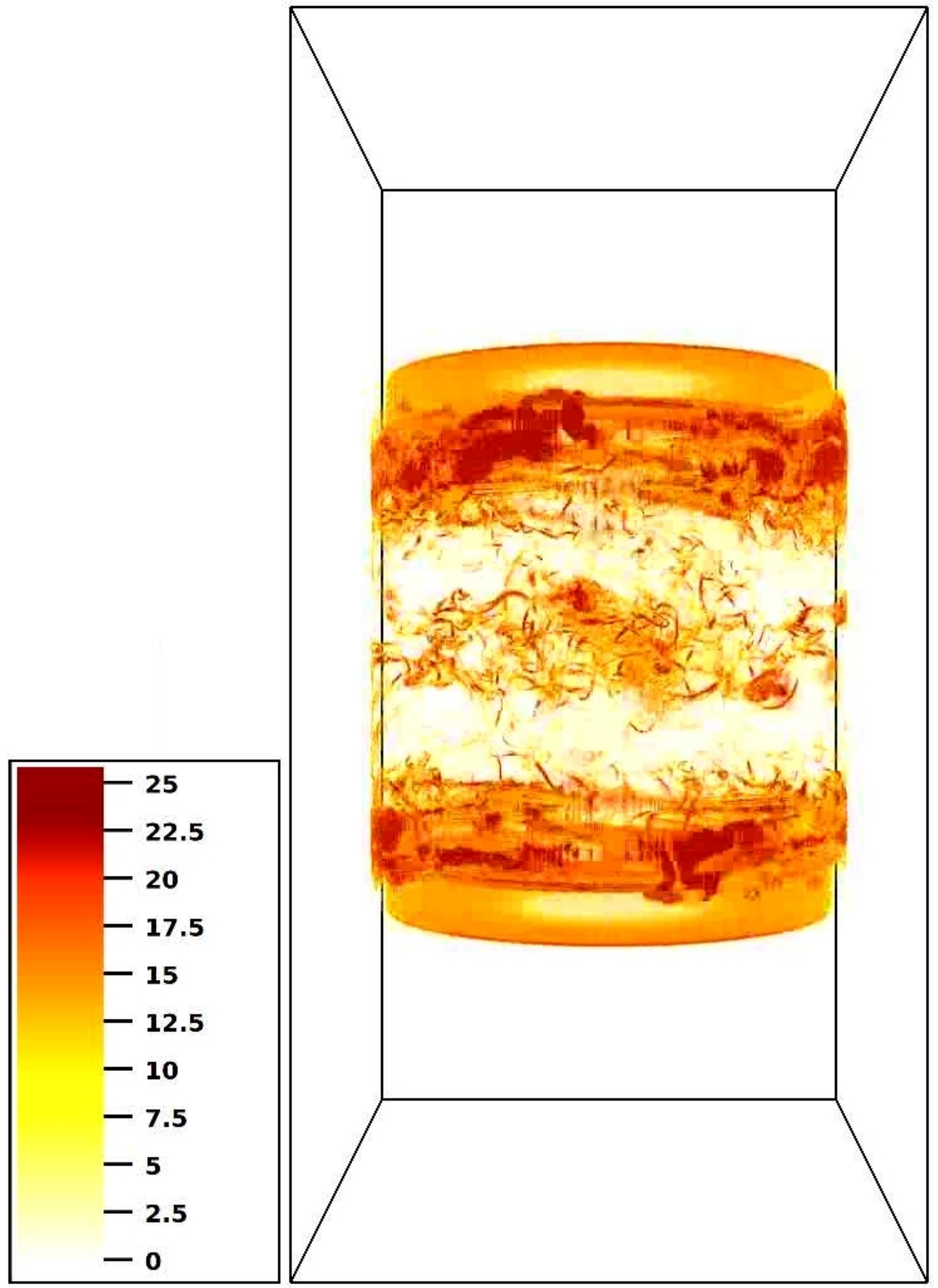}}
\put(25,2){\includegraphics[height=0.33\textwidth]{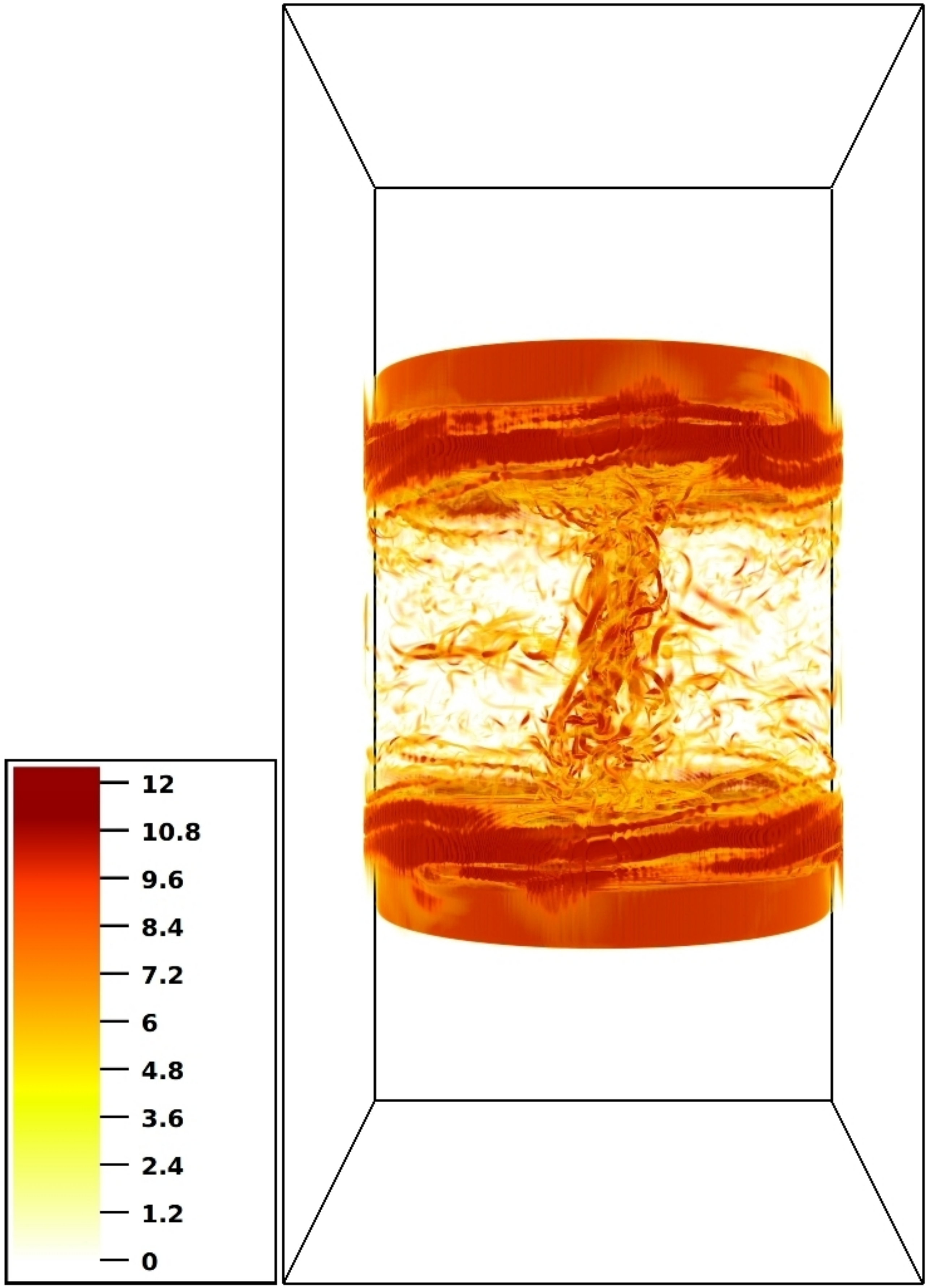}}
\put(12.7,32){Case S}
\put(37.2,32){Case C+}
\end{picture}
\caption{\label{enstr}(Color online) Volume rendering of the instantaneous enstrophy field $\mathcal{Z}=\bm{\omega}^2/2$ for the straight blades geometry (left) and for the positive-rotating curved blades one (right). The full spatial numerical domain is also delineated.}
\end{center}
\end{figure}

\begin{figure}
\begin{center}
\setlength{\unitlength}{0.01\textwidth}
\begin{picture}(48,63)
\put(0,0){\includegraphics[width=0.45\textwidth]{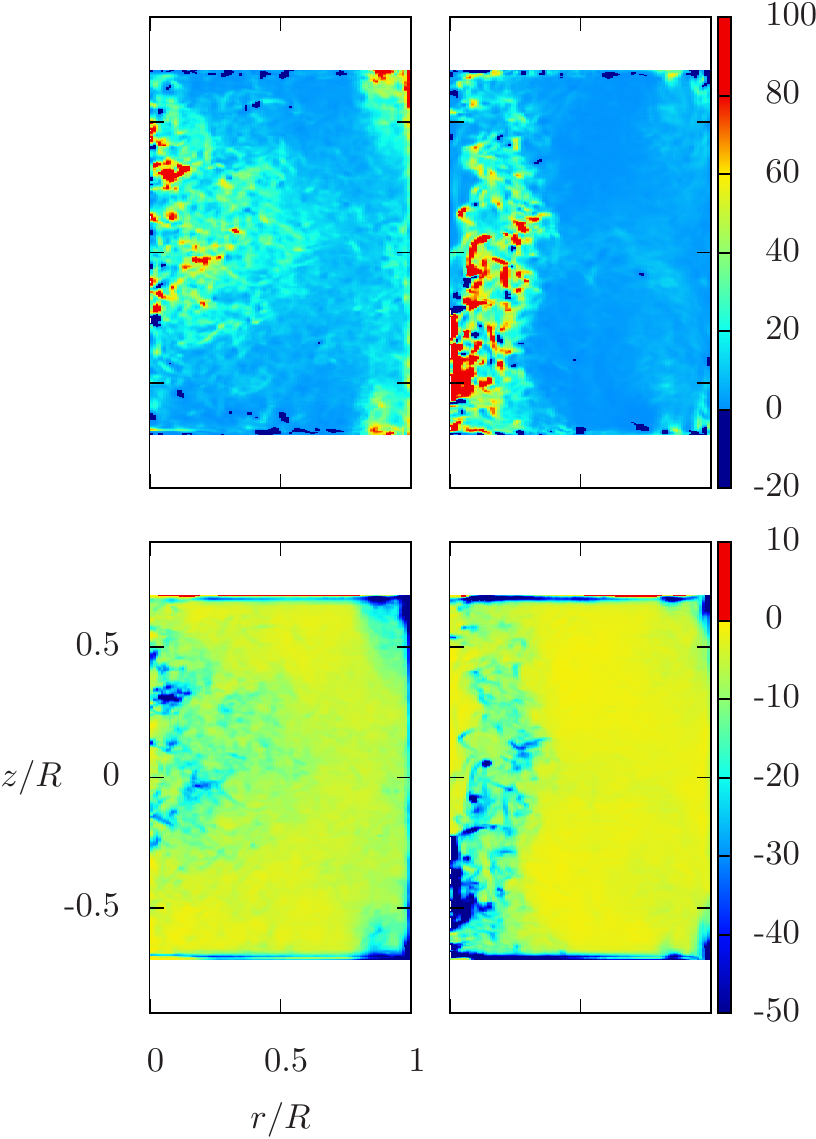}}
\put(12.8,33.5){Case S}
\put(28.8,33.5){Case C+}
\end{picture}
\caption{\label{stretch}(Color online) Instantaneous distribution of the vortex stretching term $\omega_i\omega_jS_{ij}$ (top) and the viscous term $\nu\omega_i\nabla^2\omega_i$ (bottom) of the evolution equation of the enstrophy \eqref{enstr_evol} for the straight blades (left) and the curved blades (right). Both quantities are non-dimensionalized by $\left<\mathcal{Z}\right>\left<\varepsilon\right>/k$, where $\left<\varepsilon\right>$ is the mean dissipation rate.}
\end{center}
\end{figure}

\textit{Turbulence properties} -- In the core region, the flow corresponds to inertial turbulence. The Reynolds number based on the Taylor micro-scale is $\mathrm{Re}_\lambda\approx120$ for the straight blades (case S) and $\mathrm{Re}_\lambda\approx80$ for the curved blades (case C+), although $\mathrm{Re}_K$ is unchanged. This demonstrates that the turbulent
field intensity depends importantly on the geometry of the blades at equivalent rotation frequency. The small scale turbulent structures are depicted
 on figure \ref{enstr}  with isosurfaces of the enstrophy $\mathcal{Z}=\bm{\omega}^2/2$ for cases S and C+.
These figures from DNS fields allow an original characterization of unexpected turbulence features, not accessible
to the experimental
observations:  the intense turbulent vortices are clustered around the horizontal symmetry plane at $z=0$ for the straight blades simulation, showing that most of the vorticity production is due to the horizontal shear layer of azimuthal velocity; for the curved blades flow, the distribution of enstrophy highlights  intense vortical structures along the axis of rotation and the horizontal clustering is almost not observed in this configuration. The mechanism of turbulence generation seems to  vary
strongly with the shape of the blades. We confirm this by considering the dynamical equation of enstrophy $\mathcal{Z}$
\begin{equation}
\mathrm{D}_t\mathcal{Z}=\omega_i\omega_jS_{ij}+\nu\omega_i\nabla^2\omega_i,
\label{enstr_evol}
\end{equation}
where $S_{ij}=(\partial_i u_j  +\partial_j u_i)/2$ is the rate of strain tensor and $\mathrm{D}_t=\partial_t+\bm{u}\cdot\bm{\nabla}$. The first term on the right-hand-side of equation \eqref{enstr_evol} is the vortex-stretching term and is principally responsible for enstrophy production. The second term linked to viscosity can also be a source of vorticity and enstrophy production by the tilting of vortices or by helping reconnection (see \cite{holzner2010viscous}), but this is globally not the case for these flows, as can be seen on figure \ref{stretch} (very few occurences of $\nu\omega_i\nabla^2\omega_i>0$).
%
%
%
This figure shows the vortex stretching term and the viscous term for cases S and C+:  we first observe that the straight blades produce highly turbulent small structures close to their extremities ($r^*\rightarrow1$), even when the stationary state is reached. This mechanism is different in run C+ since it is no longer observed once the mean recirculation cells are established. Second, 
with straight blades, enstrophy production is localized in the horizontal shear layer and the fluctuations are advected by the mean flow towards the center of the fluid domain. This phenomenon changes with curved blades, where the vortices are created by the large negative radial velocities close to the center and by the vertical shear induced by azimuthal velocities of opposite sign within the recirculation cells. This explains why the turbulent structures are almost confined to a region along the vertical axis of rotation. The generation of turbulence is thus very different according to the shape of the blades, which
also explains the different total energy and mean flow topology. 

\begin{figure}
\includegraphics[width=0.4\textwidth]{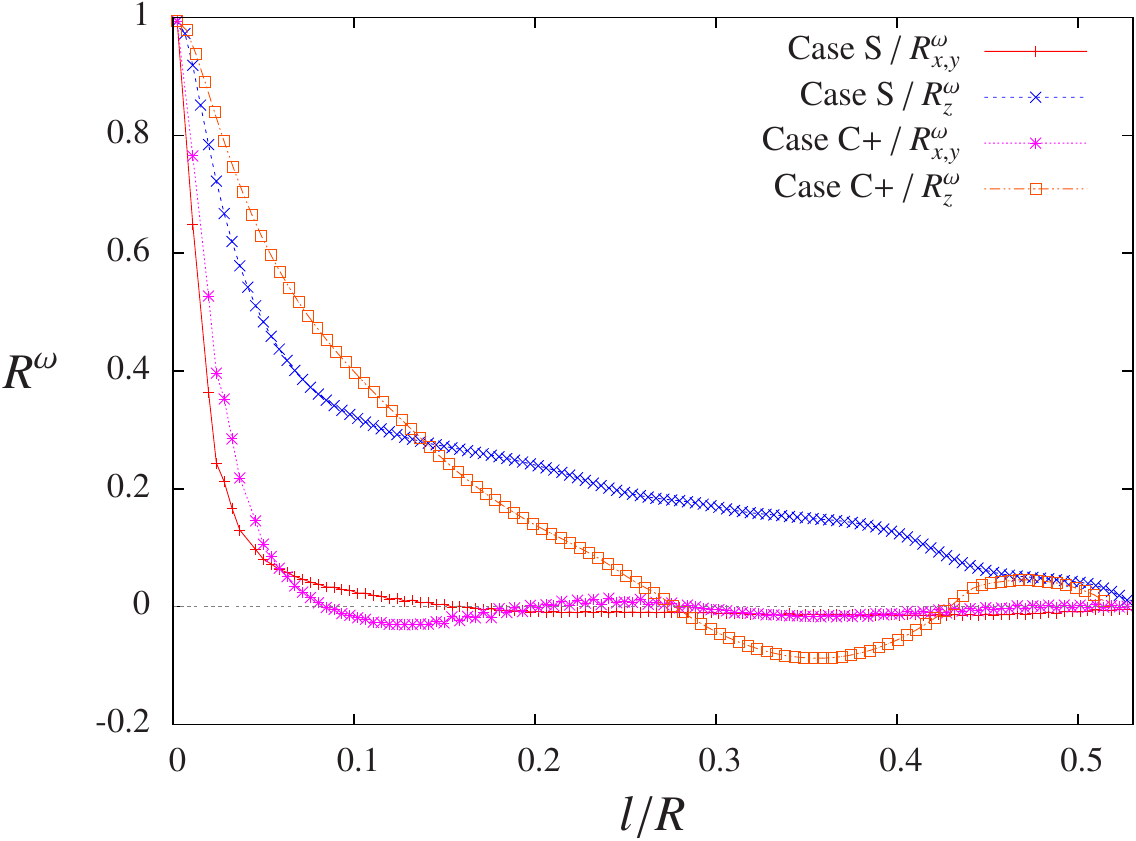}
\caption{\label{correl}(Color online) Correlation function based on the vertical vorticity component $\omega_z$ for a horizontal or a vertical separation, of the flows generated by straight (S) or curved blades (C+).}
\end{figure}

\textit{Anisotropy of turbulence} -- We now refine the characterization of turbulence in the center of the fluid domain. We define a core region $\mathcal{D}_c$ where $r^*<0.3$ and $2\vert z\vert/H<0.3$ and we compute from the S and C+ velocity fields the Taylor micro-scale in each direction defined by
\begin{equation}
\lambda_i=\left( (4k) / \left( 3 \left(\partial_i u_i\right)^2 \right) \right)^{1/2}
\end{equation}
(no summation over $i$). We previously observed that the Reynolds number $\mathrm{Re}_\lambda$ is larger in the straight blades case, but this is essentially due to a larger kinetic energy, the Taylor scales $\lambda_i$ being lower ($\lambda_z\approx0.38$ for the case S and $\lambda_z\approx0.42$ for the case C+ for example). In both cases, the vertical length scale $\lambda_z$ is greater than the horizontal ones $\lambda_{x,y}$. The difference is about $5\%$ in case S and reaches about $15\%$ in case C+. The separation of the
directional Taylor scales shows that the flow is not isotropic in the small scales, also confirmed by the plot of the two-point  vertical vorticity correlation function
\begin{equation}
R^\omega_i(l)=\frac{\left<\omega_z(\bm x)\omega_z(\bm x+l\bm e_i)\right>_{\mathcal{D}_c}}{\sqrt{\left<\omega_z^2(\bm x)\right>_{\mathcal{D}_c}\left<\omega_z^2(\bm x+l\bm{e}_i)\right>_{\mathcal{D}_c}}},
\end{equation}
where $<\cdot>_{\mathcal{D}_c}$ denotes 
 spatial averaging over $\mathcal{D}_c$. The vorticity correlation function differs from the 
 velocity one, since in the latter  the mean flow contribution, although relatively weak,  prevents to estimate
 accurate turbulent correlation lengths (one could also subtract this mean flow in the velocity field, although we have not
 done so here). Figure \ref{correl} shows that vorticity is much more correlated in the vertical direction than in the horizontal ones, which is consistent with the computed Taylor scales, the vertical ones being larger than the horizontal ones.
It also clearly emphasizes the anisotropy of turbulence in the central region,  for all blades geometry, contrarily
to some initial assumptions when experimental characterization of von K\'arm\'an swirling flows began.

\textit{Beltramization} -- We consider  the relative orientation of velocity and vorticity  (see \cite{pelz1986helical,kerr1987histograms}) by computing the relative helicity 
$h={\bm{u}\cdot\bm{\omega}}/{\vert\bm{u}\vert\vert\bm{\omega}\vert}=\cos{\theta}$,
$\theta$ being the angle between the two vectors.
This angle is also important because it  measures somehow the amplitude of the non-linear term in equation \eqref{NSpen} and  the efficiency of the energy cascade. We compute the probability density function (\pdf) of $h$ and  compare it
to  previous observations in simulations of homogeneous isotropic turbulence   \cite{rogers1987helicity,kerr1987histograms}.
\begin{figure}
\includegraphics[width=0.45\textwidth]{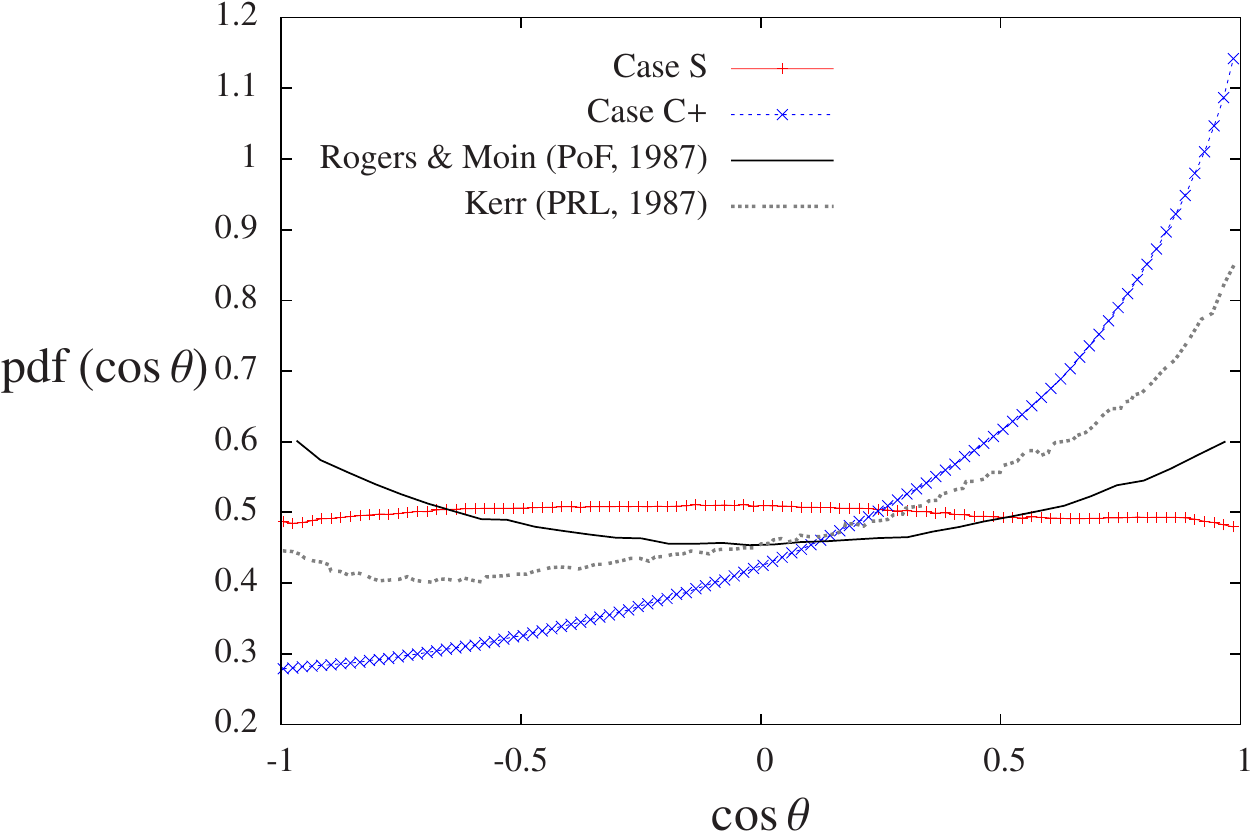}
\caption{\label{pdfhelic}(Color online) \textit{Pdf}s of  $\cos\theta$ --- where $\theta$ is the angle between velocity and vorticity vectors --- equal to the `relative helicity' $h$. The solid black line denotes the results of isotropic simulations by Rogers \& Moin \cite{rogers1987helicity} and the dotted grey one those of Kerr \cite{kerr1987histograms} with a large-scale forcing.}
\end{figure}
The results  presented on figure \ref{pdfhelic}  show surprising features. On the one hand, in the straight blades case 
S, the \pdf~of $h$ is approximately flat, as if velocity and vorticity were uncorrelated, thus differing from  the homogeneous
isotropic  turbulence \pdf, which curves up at $\cos\theta=\pm1$. On the other hand, run C+ shows that the flow enhances  velocity-vorticity alignment and inhibits anti-alignment. This is explained by the presence of the mean flow which is non-negligible even in the region $\mathcal{D}_c$, as shown by figure \ref{mean} for case C+. As the vorticity is mainly vertical (see figure \ref{enstr} with the intense vertical vortices) and more likely to be aligned with the rotation of the closer disk, and as the vertical velocity is linked to the mean recirculation cells, we believe that this alignment is not a feature of  turbulence but a trace of the forcing by the mean flow. Opposite rotation of the disks would lead to the anti-symmetric distribution with a peak at $\cos\theta=-1$. 
But we also note that a similar distribution of the \pdf~of $h$ is also observed by \cite{kerr1987histograms} in simulations of  isotropic turbulence 
with large-scale forcing. Overall, this confirms that forcing turbulence modifies in some way its `natural' dynamics and
structure, and that high Reynolds number homogeneous isotropic turbulence may be very hard to achieve.

\textit{Conclusion} -- We have  highlighted the importance of the shape of the blades on the generation of turbulence, and on the turbulence itself. Differences are observed for the mean flows even  at identical $\mathrm{Re}_K$, since
turbulence production can appear in different zones of the cell.  Visualizations, directional Taylor micro-scales and correlation functions show the non negligeable anisotropy of turbulence in the core region.

\textit{Acknowledgments} -- Computational time was provided
by CINES under project number \texttt{flu2206}.





%

\end{document}